\documentclass[aps,prd,final, nofootinbib]{revtex4}
\usepackage{amsmath, amsfonts, amsthm, amssymb, graphicx,epsfig}
\usepackage[paperwidth=210mm,paperheight=297mm,centering,hmargin=2.5cm,vmargin=2.5cm]{geometry}
\usepackage{subfig}
\usepackage{epstopdf}
\usepackage{color}
\usepackage{tikz}
\usepackage{amsthm}
\usepackage{enumerate}
\theoremstyle{plain}
\newtheorem{thm}{Theorem} 
\theoremstyle{definition}
\newtheorem{defn}[thm]{Definition} 
\def\be{\begin{equation}}
\def\ee{\end{equation}}
\def\ba{\begin{eqnarray}}
\def\ea{\end{eqnarray}}

\def\bdm{\begin{displaymath}}
\def\edm{\end{displaymath}}

\def\bq{\begin{quote}}
\def\eq{\end{quote}}

\def\d{{\rm d}}
\def\del{\partial}
\def\ltap{\ \raise.3ex\hbox{$<$\kern-.75em\lower1ex\hbox{$\sim$}}\ }
\def\gtap{\ \raise.3ex\hbox{$>$\kern-.75em\lower1ex\hbox{$\sim$}}\ }
\def\gl{\ \raise.5ex\hbox{$>$}\kern-.8em\lower.5ex\hbox{$<$}\ }
\def\roughly#1{\raise.3ex\hbox{$#1$\kern-.75em\lower1ex\hbox{$\sim$}}}

 at 11truept

\def\d2{{(\hat D \pi)^2 }}

\newcommand{\beq}{\begin{equation}}
\newcommand{\eeq}{\end{equation}}
\newcommand{\bea}{\begin{eqnarray}}
\newcommand{\eea}{\end{eqnarray}}
\newcommand{\beqa}{\begin{eqnarray}}
\newcommand{\eeqa}{\end{eqnarray}}
\newcommand{\nn}{\nonumber\\}

\def \1 {\pi_{i_1}}
\def \2 {\pi_{i_2}}
\def \3 {\pi_{i_3}}
\def \4 {\pi_{i_4}}
\def \5 {\pi_{i_5}}
\def \6 {\pi_{i_6}}
\def \7 {\pi_{i_7}}

\begin{document}
\title{Proof of the most general multiple-scalar field theory in Minkowski space-time free of Ostrogradski Ghost}
\author{Vishagan Sivanesan}
\email[]{svishagan@gmail.com}
\affiliation{School of Physics and Astronomy, University of Nottingham, Nottingham NG7 2RD, UK} 
\date{\today}

\begin{abstract}
We give a proof of the most general multiple scalar field theory in flat space-time free of Ostrogradski ghosts. We start from the assumption that the action is a functional of the scalar fields and their derivatives of order upto two (which in general can have field equations of derivative order up to four), and arrive at the most-general field theory that can be constructed such that the field equations are of derivative order up to two. Naturally, this theory includes all physically viable multi-scalar field theories in flat space-time, in the sense that they don't suffer from Ostrogradski instability \cite{ostro}.
\end{abstract}


\maketitle
\section{Introduction}
Recently, there has been a lot of interest in scalar field theories with higher derivative structure, and with derivative self-interaction. These theories have been investigated, especially in the context of modifying gravity in the infra-red. Famous examples include, the effective field theory of the DGP model \cite{dgp} on the brane, more precisely the \textit{decoupling limit} of this theory, and the generalisation of this effective theory to higher order in fields, known as, the Galileon field theory \cite{gal}\cite{tony}. These field theories with derivative self-interactions exhibit novel properties at both the classical and quantum mechanical level. The \textit{Vainshtein mechanism} \cite{hornvainsh,vainsh,deff} which suppresses the scalar force near dense environments, local strong coupling \cite{qdgp}, violation of null-energy condition without pathological ghost or gradient instabilities (see \cite{nico-early} \citep{vikman4}\cite{vikman1}\cite{null}\cite{sublum}\cite{vikman2}\cite{vikman3}\cite{dbigal}\cite{rubakov}) are a few such properties. In \cite{dgsz} the authors generalised the single Galileon field theory to find the most general scalar field theory in Minkowski space-time with field equations of derivative order up to two. Remarkably when this theory was covariantised and appropriate non-minimal couplings added, it reduced to the most general scalar-tensor theory in arbitrary background in 4-dimensions which was proved by Horndeski in 1974 \cite{horny} (see also \cite{covgal} for an earlier covariantisation of galileon field theory). Later this covariantisation procedure was carried out for the multiple scalar field theory (that is proved in this paper to be the most general one), and was conjectured to be the most general multi-scalar-tensor theory in arbitrary dimensions and arbitrary backgrounds \cite{vishmulti}. This paper can be taken to be a companion to \cite{vishmulti} where the conjectured starting point, the most general multiple scalar field theory, is proved.

\section{Statement of the proof}
The most general multiple scalar field theory in flat space-time satisfying the conditions
\begin{enumerate}[i)]
\item Lagrangian contains up to second order derivatives of the fields
\item Field equations contain up to second order derivatives of the fields
\end{enumerate}
 is given by,
\be \label{6genmgal}
S_\text{multi-scalar}= \int_{\cal M} d^D x ~A(\bar X_{ij}, \pi_l)+  \sum_{m=1}^{D-1} A^{k_1 \dots k_m} (\bar X_{ij}, \pi_l)\del^{[a_1}\del_{a_1} \pi_{k_1} \cdots \del^{a_m]}\del_{a_m} \pi_{k_m}
\ee
where $\bar X_{ij}=\frac{1}{2} \del_a \pi_{i} \del^a \pi_j$ and $\frac{\del A^{i_1 \dots i_m}}{\del \bar X_{kl}}$ is symmetric in {\it all} of its indices $i_1, \ldots i_m, k, l$. 

\section{Polynomiality of the second derivatives of $\pi_i$ in the Lagrangian}
We start with the general multi-field action in D dimensions of the form,
\be 
S = \int d^D x \mathcal{L}\left( \pi_{i}, \del_a \pi_{j}, \del_{b}\del_c \pi_{k} \right)
\ee
Here $i,j,k$ are used to label different fields $(i \in \{ 1 \dots N \})$. The Euler-Lagrange equations of this action is 
\be 
\frac{\del \mathcal{L}}{\del \pi_{i}} - \del_a \left(\frac{\del \mathcal{L}}{\del \pi_{i\,a}}\right) + \del_a \del_b \left(\frac{\del \mathcal{L}}{\del \pi_{i\,ab}}\right) = 0
\ee
Note that we occasionally resort to the notation $\pi_{i\,a \dots b} \equiv \del_b \dots \del_a \pi_i$ for brevity. In general only the third term on the LHS will have fourth derivatives, explicitly this is,
\be 
\frac{\del \mathcal{L} }{ \del \pi_{i\,cd} \del \pi_{i\,ab}} \pi_{i\,abcd}
\ee
imposing the constraint that this term vanishes we get,
\be \label{6deriv4}
\mathcal{L}^{ab|cd} \pi_{abcd}=0
\ee
where we used the notation, $\mathcal{L}^{ab|cd} \equiv \frac{\del \mathcal{L} }{ \del \pi_{i\,cd} \del \pi_{i\,ab}}$ and suppressed the internal indices for brevity. Due to the complete symmetry in the space time indices of $\pi_{abcd}$, \ref{6deriv4} implies 
\be \label{6.2}
\mathcal{L}^{(ab|cd)} =0
\ee
where ($\dots$) stands for symmetrisation. Furthermore, the following symmetries are trivial.
\be \label{6symm}
\mathcal{L}^{ab|cd} = \mathcal{L}^{cd|ab}  = \mathcal{L}^{cd|ba} = \mathcal{L}^{dc|ba} 
\ee
On account of these symmetries, \ref{6.2} is satisfied \textit{if and only if} the following cyclic identity holds
\be \label{6.3}
\mathcal{L}^{ab|cd} + \mathcal{L}^{ac|db} + \mathcal{L}^{ad|bc} =0
\ee
We define the tensor,
\be \label{6tensor}
\mathcal{L}^{a_1 b_1|\dots | a_n b_n} \equiv \frac{\del}{\del \pi_{a_n b_n}} \dots \frac{\del}{\del \pi_{a_1 b_1}} \mathcal{L}
\ee
This tensor naturally inherits the cyclic identity \ref{6.3},
\be 
\mathcal{L}^{\dots |ab| \dots |cd| \dots } +\mathcal{L}^{\dots |ac| \dots |db| \dots } + \mathcal{L}^{\dots |ad| \dots |bc| \dots } =0
\ee
Consider the (D+1)th derivative of the Lagrangian with respect to $\pi_{ab}$ that is given by the tensor \ref{6tensor} of rank (0,2D+2),
\be \label{6d+1deriv}
\mathcal{L}^{a_1 b_1 | \dots | a_D b_D | a_{D+1} b_{D+1}}
\ee
In D dimensions it is easy to see that the components of this tensor will have atleast 3 identical space-time indices. Using the symmetry \ref{6symm} any component of this tensor can be cast in the following two forms,
\ba 
F_1 &=& \mathcal{L}^{\dots |aa|ab| \dots }\nn
F_2 &=& \mathcal{L}^{\dots |ab|ac|ad| \dots}
\ea
Now using the cyclic identity on the identical indices a of $F_1$ yields,
\be 
F_1 = 0
\ee
and using the cyclic identity on the first three indices $a,b,a$ of $F_2$ and subsequently on the similar indices gives,
\be 
F_2 = \mathcal{L}^{\dots |ab|ac|ad| \dots} = -\frac{1}{2} \mathcal{L}^{\dots |aa|ad|bc| \dots } =0
\ee
Thus we have established,
\begin{thm}\label{6thmpoly}
The most general action that yields equations of motion of derivative order up to two, has the Lagrangian that depends polynomially on the second derivatives of the field and the polynomial order is bounded above by D in D-dimensions ie, $\text{Order of the polynomial} \leq D $.
\end{thm}
\section{The Structure of the Lagrangian $\mathcal{L}$}
Having established the polynomiality of the second derivatives in $\mathcal{L}$, we write down a generic term in the Lagrangian that is constrained by this fact.
\ba
\mathcal{L}_{2p,q} &=& A(\bar X_{ij}, \pi_l)^{i_1 \dots i_p j_1 \dots j_q} \sum_{\sigma \in S_{p+q}} \sum_{p \in S_q} c(\sigma) d(p) \left [ \left(\del_{a_1}\pi_{i_1} \del^{\sigma(a_1)}\pi_{i_2}\right) \dots \left(\del_{a_p} \pi_{i_{(2p-1)})} \del^{\sigma(a_p)}\pi_{i_{2p}}\right) \right. \nn
&&\qquad\qquad\qquad\qquad \left. \times \del_{p(b_1)}^{\sigma(b_1)}\pi_{j_1} \dots  \del_{p(b_q)}^{\sigma(b_q)}\pi_{j_1} \right]
\ea
A note on notations is in order. Here $i,j,k \dots $ denote internal indices and $a,b,c \dots $ are space-time indices. $\bar X_{ij}=\frac{1}{2} \del_a \pi_{i} \del^a \pi_j$ and $A(\bar X_{ij}, \pi_l)^{i_1 \dots }$ can in general be a non-polynomial function. $\sigma \in S_{p+q}$ is an element of the symmetric permutation group acting on the positions of the space-time labels $a_i, b_j$ upstairs, similarly $p \in S_q$ is an element of the symmetric permutation group acting on the positions of the space-time labels $b_j$ downstairs. $C_\sigma,d_p$ are coefficients that depend on $\sigma,p$ resp (in general they can also be functions of $\bar X_{ij},\pi_k$). 
We observe the following facts.
\begin{enumerate}[i)]
\item Any term generated by the action of $\sigma$ \textbf{within} the positions of the indices $b_i$ can also be generated via the action of $p$ on $b_i$ downstairs since the indices are summed over.
\item Any term generated by the action of $\sigma$ \textbf{within} the positions of the indices $a_i$ can also be generated by shuffling the order of internal indices $i_j$.
\end{enumerate}
On account of these facts we can avoid over-counting by reducing the symmetry group of $\sigma$ to the quotient group defined by,
\be 
S_{p+q} \to \frac{S_{p+q}}{S_p \,S_q}
\ee
\subsection{Symmetricity of $A(\bar X_{ij},\pi_k)^{i_1,i_2 \dots}$}
So far we have established that a generic term in the Lagrangian can be cast in the form,
\ba
&&\mathcal{L}_{2p,q}\left[A(\bar X_{ij}, \pi_l)^{i_1 \dots i_p j_1 \dots j_q} \right] \equiv\quad A(\bar X_{ij}, \pi_l)^{i_1 \dots i_p j_1 \dots j_q} \\
&& \times \sum_{\sigma \in \frac{S_{p+q}}{S_p \, S_q}} \sum_{p \in S_q} c(\sigma) d(p) \left [ \left(\del_{a_1}\pi_{i_1} \del^{\sigma(a_1)}\pi_{i_2}\right) \dots \left(\del_{a_p} \pi_{i_{2p-1})} \del^{\sigma(a_p)}\pi_{i_{2p}}\right) \right.
 \left.\del_{p(b_1)}^{\sigma(b_1)}\pi_{j_1} \dots  \del_{p(b_q)}^{\sigma(b_q)}\pi_{j_1} \right]\nonumber
\ea
We note that any cancellation of 3'rd and 4'th derivatives coming from the variational principle should occur within each of these terms as they are structurally different from each other. We ignore the variation of $A^{i_1\dots}$ for now, and focus on terms that are of derivatives order 3 (containing $(\pi_{i\,abc})$). Let us isolate a generic term that would give rise to 3'rd order derivatives in $\mathcal{L}_{(2p,q)}$.
\be 
\mathcal{L}_{2p,q} \supset B_{k,l} \equiv A^{\dots k \dots l \dots } \sum_{\sigma,p}c_{\sigma}d_{p} \left( \del^{\sigma(a_r)}\pi_k\,\del^{\sigma(a_s)}_{p(b_t)}\pi_l \right) \dots
\ee
where we have suppressed the dependence of the arbitrary function $A^{i\dots}(\bar X_{ij}, \pi_k)$ and other factors of the fields for brevity. For the $\delta \pi_k$ variation this will contain the term ,
\be 
\delta B_{k,l} \supset - A^{\dots k \dots l \dots }\sum_{\sigma,p} c_{\sigma}d_{p} \left( \delta \pi_k\,\del^{\sigma(a_r)}\del^{\sigma(a_s)}_{p(b_t)}\pi_l \right) \dots
\ee
Note however that under the interchange of indices $k,l$ the corresponding variation yields a similar term \textit{i.e,}
\be 
\delta B_{l,k} \supset + A^{\dots l \dots k \dots }\sum_{\sigma,p} c_{\sigma}d_{p} \left( \delta \pi_k\,\del^{\sigma(a_r)}\del^{\sigma(a_s)}_{p(b_t)}\pi_l \right) \dots
\ee
These would cancel if,
\be 
A^{\dots k \dots l \dots } = A^{\dots l \dots k \dots }
\ee
This implies that $A^{i_1 \dots}(\bar X_{ij},\pi_k)$ is completely symmetric in it's indices.

As we have seen the symmetry group of the $\sigma$ would just amount to the interchange of space-time labels $a_i \leftrightarrow b_j$. On account of the symmetry in $A^{i_1 \dots}$, it does not matter which labels are interchanged \textit{i.e,} all such operations form an \textit{equivalence class} labelled by how many pairs are interchanged. Now consider the case where two interchanges are made, we write a generic term of this type, and omit the internal indices and the function $A^{i\dots}$ as they don't play any role in this argument, due to the symmetricity of $A^{\i \dots}$,
\be 
C \equiv c_p c_{\sigma}\pi_{a_i} \pi_{a_j} \pi^{b_k} \pi^{b_l} \pi^{a_i}_{p(b_k)} \pi^{a_j}_{p(b_l)} \dots 
\ee
Here we have interchanged $ a_i \leftrightarrow b_k, a_j \leftrightarrow b_l$. This term would give rise to 4'th derivatives in the eom given by,
\be 
 \delta_{ij} C \supset 2c_p c_{\sigma} \pi_{a_i} \pi_{a_j} \pi^{b_k} \pi^{b_l} \left(\pi^{a_i}_{p(b_k)}{}^{a_j}_{p(b_l)}\right) \delta \pi \dots 
\ee
Here $\delta_{ij}$ denotes the variation and the subsequent integration by parts restricted to the factors $\pi^{a_i}_{p(b_k)} \pi^{a_j}_{p(b_l)}$. Note that $C$ is invariant under the interchange of either $a_i \leftrightarrow a_j$ or $p(b_k) \leftrightarrow p(b_l)$, this means that the 4'th derivative term in $\delta C$ would vanish only if all terms proportional to $C$ vanish. Thus we have established, 
\begin{thm}
The action of $\sigma$ is necessarily limited to just the interchange of at most 1 index $a_i \leftrightarrow b_j$, if the condition (ii) is to be satisfied. Thus the $\mathcal{L}_{2p,q}$ is restricted to
\ba
\mathcal{L}_{0,q}\left[A^{i_1 \dots i_{q}} \right] &=& A^{i_1 \dots i_{q}}\sum_{p \in S_q} c_p \del^{b_1}_{p(b_1)}\pi_{(i_1)} \dots \del^{b_q}_{p(b_q)}\pi_{(i_q)}\\
\mathcal{L}_{2,q}\left[A'^{i_1 \dots i_{q+2}}\right] &=& A'^{i_1 \dots i_{q+2}} \sum_{p \in S_q} c_p \del_{a} \pi_{i_1} \del^{b_1} \pi_{(i_2)} \del^{a}_{p(b_1)}\pi_{(i_3)} \dots \del^{b_q}_{p(b_q)}\pi_{(i_{q+2})}
\ea
\end{thm}

Having resolved the $\sigma$ permutations we focus on the permutation $p$ acting on the indices $b_i$ downstairs. 
\begin{defn}
$T^{ij}$ is a transposition map that acts on individual instances of the permutations generated by $p$ and interchanges $b_i \leftrightarrow b_j$, 
\be\nonumber
T^{ij}: P_k^{ij} \to P_k^{ji}
\ee
Note that $T^{ij}\circ T^{ji} = id$. Here $ P_k^{ij}$ is an instance of the permutation where $b_i$ appears before $b_j$ and $k \in \{1,\dots q!/2\}$.
\end{defn}
Two permutations $P_k^{ij},P_{k'}^{i'j'}$ are identical if they are related by the relabelling of the indices $b_i$. Relabelling of $b_k \leftrightarrow b_l$ would change their order both upstairs and downstairs, which can be put back into the allowed form by interchanging the fields that carry the relabelled indices upstairs. Let us illustrate this by an example. Consider the following term in the series, where we have again omitted the symmetric function $A^{i_1 \dots}$ and the remaining factors of the field,
\be \label{6ex1}
\del_{b_4}^{b_2}\pi_{i_r}   \del_{b_6}^{b_3} \pi_{i_{r+1}} \dots \del_{b_2}^{b_7}\pi_{i_{r+5}}  \del_{b_3}^{b_8}\pi_{i_{r+6}}
\ee
relabelling $b_2 \leftrightarrow b_3$ yields the identical term,
\be 
\del_{b_4}^{b_3}\pi_{i_r}   \del_{b_6}^{b_2} \pi_{i_{r+1}} \dots \del_{b_3}^{b_7}\pi_{i_{r+5}}  \del_{b_2}^{b_8}\pi_{i_{r+6}}
\ee
But permutation group $p$ would not alter the indices upstairs therefore we have to move the fields back such that the order upstairs is unchanged. We can interchange the positions of the terms $\del_{b_4}^{b_3}\pi_{i_r} ,  \del_{b_6}^{b_2} \pi_{i_{r+1}}$ invariantly since the function $A^{i_1 \dots}$ is symmetric. We get the identical term
\be 
\del_{b_6}^{b_2} \pi_{i_{r+1}} \del_{b_4}^{b_3}\pi_{i_r} \dots \del_{b_3}^{b_7}\pi_{i_{r+5}}  \del_{b_2}^{b_8}\pi_{i_{r+6}}
\ee
comparing this to \ref{6ex1} we see that it differs by two transpositions acting on the original term interchanging $b_2 \leftrightarrow b_3$ and $b_4 \leftrightarrow b_6$, in other words it is operated by the map $T^{23} \circ T^{46}$. Thus we have established,
\begin{thm}\label{6thmdist}
Two identical terms generated by the permutation group $p$ \textbf{necessarily} differ by an even number of transpositions acting downstairs. 
\end{thm}

\begin{defn}
$M^{[ik][jl]}$ is a map acting on the permutation $P_r^{ij}$ and relabels $b_i \leftrightarrow b_k$ and $b_j \leftrightarrow b_l$ i.e,
\be 
M^{[ik][jl]} :  P_r^{ij} \to  P_r^{kl}
\ee
The two maps $T^{ij}$ and $M^{[ik][jl]}$ can be expressed succinctly with the above commutative diagram.
\begin{figure}[h!]\label{6commdiag}
\centering
\includegraphics{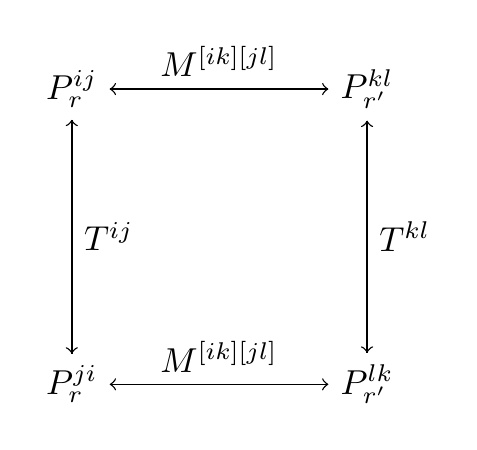}
\caption{Commutative diagram of the maps $T^{ij}, M^{[ik][jl]}$}
\label{6commdiag}
\end{figure}
It is clear from this diagram that there is an induced map $M^{[ik][jl]} : T^{ij} \to T^{kl}$.
\end{defn}

Consider a given term in the permutation series, where once again we drop the internal indices and focus only on the two relevant factors for brevity,
\be 
c_p P_k^{ij} =  c_p \dots \del^{b_{r}}_{b_i}\pi \dots \del^{b_s}_{b_j}\pi \dots 
\ee
Under the action of $T^{ij}$ we get,
\be 
c_{p'} P_k^{ji} = c_{p'} \dots \del^{b_{r}}_{b_j}\pi \dots \del^{b_s}_{b_i}\pi \dots 
\ee
Both of these terms are distinct on account of theorem \ref{6thmdist}. Varying the second derivatives and integrating by parts would yield for each of these terms
\be 
\delta_{ij}\left(c_p P_k^{ij}\right)\supset 2c_p \dots \delta\pi \dots \del^{b_{r}}_{b_j}{}\del^{b_s}_{b_i}\pi \dots 
\ee
and
\be 
\delta_{ji}\left(c_{p'} P_k^{ji}\right) \supset 2c_{p'} \dots \delta\pi \dots \del^{b_{r}}_{b_j}{}\del^{b_s}_{b_i}\pi \dots 
\ee
Note that there is also a natural induced map between the coefficients, $T^{ij}_{\text{induced}}:c_p \to c_{p'}$. Now let us observe the following facts,
\begin{enumerate}[i)]
\item Every 4'th derivative term occurring in the field equation originates from the pairs $(P_k^{ij},P_k^{ji}),(P_{k'}^{kl},P_{k'}^{lk})$ connected by transposition map and the map $M^{[ki][lj]}$ as shown in fig(\ref{6commdiag}).
\item We can set the coefficients $c_p$ of terms in Lagrangian differing by relabelling of the indices(i.e same terms) to be equal to each other without loss of generality. 
\end{enumerate}
This implies that under the action of any transposition map $T^{ij}$ we must have $T^{ij}_{\text{induced}}:c_p \to -c_p$, if the 4'th derivative terms are to cancel. At the level of Lagrangian this implies that, interchanging any indices $b_i$ downstairs would flip the sign of the Lagrangian. Thus we have resolved the permutation coefficients $c_p$, that is to say, all the indices $b_i$ downstairs are anti-symmetrised.
Now we can express $\mathcal{L}_{2,q}, \mathcal{L}_{q}$ as,
\ba 
\mathcal{L}_{2,q}\left[A^{i_1 i_2 \dots i_{q+2}} \right] &=& A^{i_1 i_2 \dots i_{q+2}}\, \del_a \pi_{i_1}\, \del^{b_1} \pi_{i_2} \,\del_{[b_1}^{a}\pi_{i_3}\, \del^{b_2}_{b_2} \pi_{i_4} \dots \del^{b_q}_{b_q]} \pi_{i_{q+2}}\\
\mathcal{L}_{q}\left[ A'^{i_1 i_2 \dots i_{q}}\right] &=& A'^{i_1 i_2 \dots i_{q}}\,\del_{[b_1}^{b_1}\pi_{i_1}\, \del^{b_2}_{b_2} \pi_{i_2} \dots \del^{b_q}_{b_q]} \pi_{i_{q}}
\ea
In order for the 3'rd derivative terms coming from the variation of the function $A,A'$ to cancel, we impose the following condition, $\frac{\del A^{i_1 \dots i_m}}{\del \bar X_{kl}}$ is symmetric in {\it all} of its indices $i_1, \ldots i_m, k, l$. 

Finally, generalising to multiple fields, a recursive relation shown in \cite{dgsz}, we find,
\be \label{recurs1}
\mathcal{L}_{2,q}\left[ G^{i_1 \dots i_{n+2}} \right] = -\mathcal{L}_{0,q}\left[B^{i_1 \dots i_n} \right] -2\mathcal{L}_{0,q-1}\left[ \frac{\del B^{i_1 \dots i_n}}{\del \pi_k}\bar  X_{k i_1}\right] + (n-1) \mathcal{L}_{2,q-1}\left[ \frac{\del B^{i_1 \dots i_n}}{ \del \pi_k} \right] 
\ee
Where $B^{i_1 \dots i_n}$ is defined by a line integral along an arbitrary curve in the $Y_{ij}$ space ending at point taking values in $\bar X_{ij}'s$. Namely,
\be 
B^{i_1 \dots i_n} = \int^{\bar X_{ij}}  G^{i_1 \dots i_n k l}(Y_{mn},\pi_r) dY_{kl}
\ee
On account of this recursive relationship we can see that $\mathcal{L}_{2,q}$ is in fact a linear combination of $\mathcal{L}_{k}$ with $k\in \{0\dots q\}$. Further, for $q=D$ we have,
\be \label{recurs2}
\mathcal{L}_{0,D}\left[ A^{i_1 \dots i_{D+2}} \bar X_{i_1 i_2} \right] = \frac{D}{2} \mathcal{L}_{2,D} \left[ A^{i_1 \dots i_{D+2}} \right]
\ee
Thus we can see from \ref{recurs1}, \ref{recurs2} that $\mathcal{L}_{0,D}$ is a linear combination of $\mathcal{L}_{0,k}$ with $k \in \{0\dots (D-1)\}$. Hence, we have established our final result,
\begin{thm}
The most general multiple scalar field theory in flat space-time satisfying conditions (i), (ii) is given by,
\be \label{6genmgal}\nonumber
S_\text{multi-scalar}= \int_{\cal M} d^D x ~A(\bar X_{ij}, \pi_l)+  \sum_{m=1}^{D-1} A^{k_1 \dots k_m} (\bar X_{ij}, \pi_l)\del^{[a_1}\del_{a_1} \pi_{k_1} \cdots \del^{a_m]}\del_{a_m} \pi_{k_m}
\ee
where $\bar X_{ij}=\frac{1}{2} \del_a \pi_{i} \del^a \pi_j$ and $\frac{\del A^{i_1 \dots i_m}}{\del \bar X_{kl}}$ is symmetric in {\it all} of its indices $i_1, \ldots i_m, k, l$. 
\end{thm}


\begin{thebibliography}{99}
\bibitem{dgp}
  G.~R.~Dvali, G.~Gabadadze and M.~Porrati,
  Phys.\ Lett.\ B {\bf 485} (2000) 208
  [hep-th/0005016].
\bibitem{qdgp}
  A.~Nicolis and R.~Rattazzi,
  JHEP {\bf 0406} (2004) 059
  [hep-th/0404159].
\bibitem{nico-early}
  P.~Creminelli, A.~Nicolis and E.~Trincherini,
  JCAP {\bf 1011} (2010) 021
  [arXiv:1007.0027 [hep-th]].
\bibitem{vikman1}
  C.~Deffayet, O.~Pujolas, I.~Sawicki and A.~Vikman,
  JCAP {\bf 1010} (2010) 026
  [arXiv:1008.0048 [hep-th]].
\bibitem{sublum}
  P.~Creminelli, K.~Hinterbichler, J.~Khoury, A.~Nicolis and E.~Trincherini,
  JHEP {\bf 1302} (2013) 006
  [arXiv:1209.3768 [hep-th]].
\bibitem{vikman2}
  I.~Sawicki and A.~Vikman,
  Phys.\ Rev.\ D {\bf 87} (2013) 6,  067301
  [arXiv:1209.2961 [astro-ph.CO]].
\bibitem{vikman3}
  D.~A.~Easson, I.~Sawicki and A.~Vikman,
  JCAP {\bf 1307} (2013) 014
  [arXiv:1304.3903 [hep-th], arXiv:1304.3903].
\bibitem{vikman4}
  D.~A.~Easson, I.~Sawicki and A.~Vikman,
  JCAP {\bf 1111} (2011) 021
  [arXiv:1109.1047 [hep-th]].
\bibitem{dbigal}
  K.~Hinterbichler, A.~Joyce, J.~Khoury and G.~E.~J.~Miller,
  arXiv:1212.3607 [hep-th].
\bibitem{rubakov}
  V.~A.~Rubakov,
  arXiv:1305.2614 [hep-th].
\bibitem{vishmulti}
  A.~Padilla and V.~Sivanesan,
  JHEP {\bf 1304} (2013) 032
  [arXiv:1210.4026 [gr-qc]].
\bibitem{deff}
C.~Deffayet and D.~A.Steer,
arXiv:1307.2450 [hep-th].
\bibitem{horny}
G.~W.~Horndeski,
Int.\ J.\ Theor.\ Phys.\  {\bf 10 } (1974)  363-384.
\bibitem{covgal}
  C.~Deffayet, G.~Esposito-Farese and A.~Vikman,
  Phys.\ Rev.\ D {\bf 79} (2009) 084003
  [arXiv:0901.1314 [hep-th]].
\bibitem{dgsz}
C.~Deffayet, X.~Gao, D.~A.~Steer and G.~Zahariade,
Phys.\ Rev.\ D {\bf 84} (2011) 064039
[arXiv:1103.3260 [hep-th]].
\bibitem{ostro} 
M. ~Ostrogradsky, 
Memoires de lÕAcademie Imperiale des Science de Saint-Petersbourg, 4:385, 
1850. 
\bibitem{gal}
A.~Nicolis, R.~Rattazzi and E.~Trincherini,
Phys.\ Rev.\ D {\bf 79} (2009) 064036
[arXiv:0811.2197 [hep-th]].
\bibitem{hornvainsh}
  A.~De Felice, R.~Kase and S.~Tsujikawa,
  Phys.\ Rev.\ D {\bf 85} (2012) 044059
  [arXiv:1111.5090 [gr-qc]].
  R.~Kimura, T.~Kobayashi and K.~Yamamoto,
  Phys.\ Rev.\ D {\bf 85} (2012) 024023
  [arXiv:1111.6749 [astro-ph.CO]].
\bibitem{vainsh}
  A.~I.~Vainshtein,
  Phys.\ Lett.\ B {\bf 39} (1972) 393.
\bibitem{null}
  A.~Nicolis, R.~Rattazzi and E.~Trincherini,
  JHEP {\bf 1005} (2010) 095
   [Erratum-ibid.\  {\bf 1111} (2011) 128]
  [arXiv:0912.4258 [hep-th]].
  A.~Masoumi and X.~Xiao,
  arXiv:1201.3132 [hep-th].
  S.~-Y.~Zhou,
  Phys.\ Rev.\ D {\bf 85} (2012) 104005
  [arXiv:1202.5769 [hep-th]].
\bibitem{tony}
  A.~Padilla, P.~M.~Saffin and S.~Y.~Zhou,
  JHEP {\bf 1012} (2010) 031
  [arXiv:1007.5424 [hep-th]].
  A.~Padilla, P.~M.~Saffin and S.~Y.~Zhou,
  Phys.\ Rev.\ D {\bf 83} (2011) 045009
  [arXiv:1008.0745 [hep-th]].
  M.~Trodden and K.~Hinterbichler,
  Class.\ Quant.\ Grav.\  {\bf 28} (2011) 204003
  [arXiv:1104.2088 [hep-th]].
\end{thebibliography}
\end{document}